\documentstyle[aps]{revtex}

\begin{document}
\title{Polytropic Stars in Three-Dimensional Spacetime}
\author{Paulo M. S\'a}
\address{\'Area Departamental de F\'{\i}sica,
U. C. E. H., Universidade do Algarve, \\
Campus de Gambelas,
8000 Faro, Portugal.
}
\maketitle
\begin{abstract}
We investigate three-dimensional perfect fluid stars with
polytropic equation of state, matched to the exterior
three-dimensional black hole geometry of Ba\~nados, Teitelboim
and Zanelli.
A new class of exact solutions for a generic polytropic index is
found and analysed.
\end{abstract}

\vskip 1cm

In the past decade, lower-dimensional theories of gravity have been 
intensively investigated, primarily because of the insights they
can provide into conceptual issues arising in four-dimensional
general relativity.
The interest in lower-dimensional theories of gravity
has dramatically increased with the discovery \cite{btz,bhtz} that
three-dimensional general relativity with a negative cosmological
constant admits a black hole solution (called BTZ), which,
apart from being asymptotically anti-de Sitter,
exhibits features closely related to the
Schwarzschild and Kerr solutions.
Further generalizations have shown that three-dimensional theories
of gravity have a very rich structure of black hole
solutions \cite{kal,hor,chan1,carl,chan2,sa1,cad,lem,sa2}.

The BTZ black hole metric can be matched to a three-dimensional
perfect fluid star.
Cruz e Zanelli \cite{cruz}, studying the stability of these
stars for the case of a generic equation of state $p=p(\rho)$,
concluded that they can undergo gravitational collapse
into a black hole.
For the particular case of pressureless dust,
Mann and Ross \cite{mann} have investigated the circunstances
under which a BTZ black hole can be formed from gravitational
collapse, concluding that it occurs naturally, 
in a way that parallels the Oppenheimer-Snyder collapse
in four dimensions.

In this letter, we consider three-dimensional perfect
fluid stars in hydrostatic equilibrium, matched to the
BTZ black hole exterior geometry.
The equation of state relating the pressure and the energy
density is assumed to be of the polytropic type,
\begin{equation}
p=K \rho^{1+\frac{1}{n}},
 \label{pes}
\end{equation}
where $n$ is the polytropic index and $K$ is the polytropic
constant.
This equation of state encompasses a wide range of physically
interesting cases, namely constant energy density for $n=0$,
nonrelativistic degenerate fermions for $n=1$,
pure radiation for $n=\infty$ and $K=1/2$,
and stiff matter for $n=\infty$ and $K=1$.
For constant energy density an exact solution
is known \cite{cruz}.
Here, exact solutions for a generic polytropic index
are found and analysed.
Note that in four spacetime dimensions analytical solutions
for polytropic stars are known just for a few particular values
of the polytropic index and only in the Newtonian approximation.

For the three-dimensional spherically symmetric, static metric,
\begin{equation}
ds^2=- e^{2\nu(r)}dt^2+e^{2\lambda(r)} dr^2+r^2 d\phi^2,
\end{equation}
the Einstein equations with a negative cosmological constant
$\Lambda<0$ in the presence of a perfect fluid
with energy density $\rho$ and pressure $p$ related by a
polytropic equation of state,
\begin{equation}
 R_{\mu\nu}-\frac12 g_{\mu\nu}R+\Lambda g_{\mu\nu}
= \kappa (\rho+p) u_{\mu}u_{\nu}+ \kappa pg_{\mu\nu},
\end{equation}
yield, after some rearrangements
\begin{eqnarray}
& & \frac{d\lambda}{dr}=\left( \kappa \rho + \Lambda \right)
                        r e^{2\lambda},
\\
& & \frac{d\nu}{dr}=\left( \kappa K \rho^{\frac{n+1}{n}}
                                           - \Lambda \right)
                        r e^{2\lambda},
\\
& & \frac{d\rho}{dr}=\frac{n}{K(n+1)} \rho^{\frac{n-1}{n}}
\left( \Lambda - \kappa K \rho^{\frac{n+1}{n}} \right)
              \left( 1 + K \rho^{\frac{1}{n}} \right)
                        r e^{2\lambda},
\end{eqnarray}
where $\kappa=2\pi G$ and $G$ is the gravitational constant.

The solution of this system of equations
for arbitrary positive polytropic index $n$ is
\begin{eqnarray}
& & e^{2\nu} = \frac{C_1}
{\left( 1 + K \rho^{\frac{1}{n}} \right)^{2(n+1)}},
      \label{nu}
\\
& & e^{2\lambda} = C_2 \frac{\left( 1+ K\rho^{\frac{1}{n}}
                             \right)^{2(n+1)}}
{\left( \kappa K \rho^{\frac{n+1}{n}} -\Lambda \right)^2},
      \label{mu}
\\
& & r^2=\frac{\Lambda}{C_2}
\left[ {\cal G}(\rho_c)-{\cal G}(\rho) \right]
+\frac{\kappa K^2}{C_2}
\left[ {\cal F}(\rho_c)-{\cal F}(\rho) \right],
      \label{r2}
\end{eqnarray}
where the following notation was introduced:
\begin{eqnarray}
& & {\cal G}(x)=\frac{1}{\left( 1+K x^{\frac{1}{n}}
                         \right)^{2(n+1)}},
\\
& & {\cal F}(x)= \frac{2(n+1)}{(n+2)}
    \frac{x^{\frac{n+2}{n}} {}_2\!F_1
    \left(1,-n;n+3;-K x^{\frac{1}{n}}\right)}
             {\left(1+K x^{\frac{1}{n}}\right)^{2(n+1)}}.
    \label{notation}
\end{eqnarray}
In the previous expressions $\rho_c$ denotes the energy density
at the center of the star and ${}_2\!F_1 (a,b;c;z)$ is the Gauss
hypergeometric function.

For positive integer and half-integer values of the polytropic
index, the hypergeometric function in (\ref{notation})
reduces to a polynomial in $x^{1/n}$.
For other values of the polytropic index $n$,
the hypergeometric function in (\ref{notation}) converges
absolutely for $K x^{\frac{1}{n}}<1$.
This condition also guarantees that in the fluid the velocity
of sound is less than the speed of light.

Solution (\ref{nu})--(\ref{r2}) fulfils the central boundary conditions
$\rho(r=0)=\rho_c$ and $\frac{d\rho}{dr}(r=0)=0$.
The surface of the star is defined by the value $r=R$ for which
$\rho=0$.
The integration constants $C_1$ and $C_2$,
which have to be positive in order to preserve the metric signature,
are determined by matching, at the surface of the star,
the interior metric (\ref{nu},\ref{mu}) to
the exterior BTZ static black hole metric
\begin{equation}
ds^2=-\left( -M_0 -\Lambda r^2 \right) dt^2
     +\frac{dr^2}{-M_0 -\Lambda r^2} 
     +r^2 d\phi^2,
   \label{mb}
\end{equation}
where the parameter $M_0$ is the conserved charge associated with
asymptotic invariance under time displacements \cite{btz}. 
Thus, one obtains
\begin{eqnarray}
& & C_1=-M_0-\Lambda R^2,
  \label{constant1}
\\
& & C_2=\frac{\Lambda^2}{-M_0-\Lambda R^2}.
  \label{constant2}
\end{eqnarray}
  
The density energy has its maximum value at the center
of the star and monotonally decreases,
reaching the value zero at the surface.
One deduces readily from (\ref{r2}) that the radius of the
star is given by
\begin{equation}
R=\sqrt{-\frac{M_0}{\Lambda}}
\left[
1-\frac{1}{{\cal G}(\rho_c)
           +\frac{\kappa K^2}{\Lambda}{\cal F}(\rho_c)}
\right]^{\frac{1}{2}}.
  \label{raio}
\end{equation}
This radius is finite for finite polytropic index $n$.

The mass of the star, given by
\begin{eqnarray}
 M &=& \int\limits_0^R 2\pi\rho(r) rdr
   \nonumber
\\
   &=& -\frac{M_0}{\Lambda^2}
       \frac{\pi K \rho_c^{\frac{n+1}{n}}
       \left( \kappa K \rho_c^{\frac{n+1}{n}} -2\Lambda \right)
              {\cal G}(\rho_c)
      -\pi\Lambda K^2 {\cal F}(\rho_c)}{{\cal G}(\rho_c)
                    +\frac{\kappa K^2}{\Lambda}
                    {\cal F}(\rho_c)},
   \label{massa}
\end{eqnarray}
is always positive;
it is also finite for finite polytropic index $n$.

The solutions described above correspond to polytropic stars of
positive index $n$ with finite mass and finite radius in
an asymptotically anti-de~Sitter spacetime. 
To construct a particular polytropic star for given values of $n$,
$\Lambda$, $M$ and $R$, we choose $M_0$ such that it obeys the
relation $-M_0-\Lambda R^2>0$ and determine the polytropic constant
$K$ and the central energy density $\rho_c$
from equations (\ref{raio}) and (\ref{massa}).
Knowing $\rho_c$ and $K$ allows
for the determination of the central pressure $p_c$, using
equation (\ref{pes}).
The whole structure of the star is thus determined.

It is know \cite{cor1,cor2} that all hydrostatic structures
with a polytropic equation of state
in three-dimensional general relativity with a vanishing
cosmological constant represent static
cosmologies; the spacetime has finite spatial volume
and cannot be matched to an exterior vacuum geometry.
The introduction of a negative cosmological constant
drastically changes this picture, allowing for
stellar solutions with finite mass and finite radius,
which match to an exterior black hole geometry
in a way similar to the situation in four dimensions.

\acknowledgments

This work was supported in part by
{\it Funda\c c\~ao para a Ci\^encia e a Tecnologia},
Portugal.

\end{document}